\documentclass[onecolumn,aps,prd,preprintnumbers,superscriptaddress,nofootinbib,amsmath,amssymb,floats,floatfix,showkeys,notitlepage,showpacs]{revtex4-1}

\usepackage{orcidlink}
\usepackage{comment}
\usepackage{lipsum}
\usepackage{graphicx}
\usepackage{subfigure}
\usepackage{palatino}
\usepackage{float}
\usepackage{sans}
\usepackage{hyperref}
\hypersetup{colorlinks=true,linkcolor=blue,urlcolor=blue,citecolor=blue}
\usepackage[toc,page]{appendix}
\usepackage[normalem]{ulem}
\usepackage{adjustbox}
\usepackage{latexsym}
\usepackage{amsmath}
\usepackage{breqn}
\usepackage{amssymb}
\usepackage{amsfonts}
\usepackage{dcolumn}
\usepackage{bm}
\usepackage{tikz}
\usetikzlibrary{shapes.geometric, arrows, positioning, arrows.meta}
\usepackage{pgfplots}
\pgfplotsset{compat=1.18}
\usepackage{bigints}
\usepackage{array,tabularx,multirow,booktabs}
\usepackage[tracking=true]{microtype}
\SetTracking{}{500}
\SetTracking{encoding={*}, shape=sc}{40}
\UseRawInputEncoding 
\allowdisplaybreaks

\begin{document}
	\newcommand \nn{\nonumber}
	\newcommand \fc{\frac}
	\newcommand \lt{\left}
	\newcommand \rt{\right}
	\newcommand \pd{\partial}
	\newcommand \e{\text{e}}
	\newcommand \hmn{h_{\mu\nu}}
	\newcommand{\PR}[1]{\ensuremath{\left[#1\right]}} 
	\newcommand{\PC}[1]{\ensuremath{\left(#1\right)}} 
	\newcommand{\PX}[1]{\ensuremath{\left\lbrace#1\right\rbrace}} 
	\newcommand{\BR}[1]{\ensuremath{\left\langle#1\right\vert}} 
	\newcommand{\KT}[1]{\ensuremath{\left\vert#1\right\rangle}} 
	\newcommand{\MD}[1]{\ensuremath{\left\vert#1\right\vert}} 

	

%
\title{\textcolor{blue}{Quasinormal modes of a dyonic black hole in Einstein\textminus Euler\textminus Heisenberg
theory }}
\author{
Shubham Kala \orcidlink{0000-0003-2379-0204}}
\email{shubhamkala871@gmail.com}
\affiliation{The Institute of Mathematical Sciences$,$ C$.$I$.$T$.$ Campus$,$ Taramani$,$ Chennai\textminus600113$,$ India}

\author{Abhishek Negi}
\email{spacetime122002@gmail.com}
\affiliation{Department of Physics$,$ Gurukula Kangri (Deemed to be University)$,$ Haridwar$,$ 249404$,$ India}

\author{Hemwati Nandan \orcidlink{0000-0002-1183-4727}}
\email{hnandan@associates.iucaa.in}
\affiliation{Department of Physics$,$ Hemvati Nandan Bahugunga Garhwal Central University$,$ Srinagar\textminus 246174$,$ Uttarakhand$,$ India\newline
Center for Space Research North\textminus West University$,$ Potchefstroom\textminus 2520$,$ South Africa}

\begin{abstract}
In this study, we investigate the quasinormal modes of a non-rotating dyonic black hole within the framework of Einstein-Euler-Heisenberg theory. We present a detailed analysis focuses on understanding the influence of dyonic charges on the oscillatory properties of these BHs. The quasinormal modes are calculated to explore the interplay between the dyonic charge and the characteristic frequencies of perturbations. The results are then systematically compared with those of black holes possessing purely electric or purely magnetic charges in the Einstein-Euler-Heisenberg framework. This comparison highlights the unique signatures and dynamic behavior introduced by the presence of dyonic charges, offering deeper insights into the properties of black holes in nonlinear electrodynamics theories.
\end{abstract}

\maketitle
\section{Introduction}
The study of black holes (BHs) in the context of extended theories of gravity has revealed a rich tapestry of solutions and phenomena that challenge our understanding of spacetime and fundamental forces \cite{born1934foundations,garcia1984type}. Among the various theoretical frameworks, Euler-Einstein-Heisenberg (EEH) theory stands out as a significant extension of general relativity (GR), incorporating nonlinear electromagnetic fields and offering insight into the interplay between gravity and electromagnetism \cite{Boillat:1970gw,Ayon-Beato:1998hmi,Ayon-Beato:1999kuh}. This theory builds upon the classical Einstein-Maxwell framework by including higher-order corrections to the electromagnetic Lagrangian, specifically designed to account for the effects of strong electromagnetic fields and quantum corrections.
In the EEH framework, the dynamics of charged BHs are governed by an extended field equation that includes terms from the Euler-Heisenberg Lagrangian \cite{Ruffini:2013hia}. This modification accounts for nonlinear corrections to the electromagnetic field, which become significant in the strong-field regime. A static and spherically symmetric BH coupled to nonlinear electrodynamics (NLED) in the weak-field limit of the Euler-Heisenberg effective Lagrangian has been extensively studied \cite{Yajima:2000kw}. This framework arises as a low-energy limit of Born-Infeld (BI) theory, revealing intriguing modifications to BH properties. Various efforts have been made to obtain regular, singularity-free BH solutions within gravitating NLED, leading to the discovery of solutions with unconventional characteristics \cite{Burinskii:2002pz,Dymnikova:2004zc,CiriloLombardo:2009zzb}. These solutions exhibit distinct features that challenge traditional expectations, offering new insights into the interplay between gravity and nonlinear electromagnetic fields \cite{Bronnikov:2000vy}. As a result, the EEH theory predicts deviations from the predictions of standard GR and Maxwell electromagnetism, leading to potentially observable effects. These include alterations in BH orbital dynamics, particle motion near the event horizon, thermodynamic properties, accretion processes, quantum electrodynamics (QED) corrections, and modifications to the BH shadow radius \cite{Amaro:2023ull,Zeng:2022pvb,Dai:2022mko,Feng:2022otu,Maceda:2018zim,Maceda:2020rpv,Rehman:2023hro,Mushtaq:2024cap}.\\
Quasinormal modes (QNMs) of BHs represent the characteristic oscillations of spacetime geometry caused by perturbations, with frequencies and damping rates determined by BH parameters such as mass, charge, and spin \cite{Hod:1998vk,Dreyer:2002vy,nollert1999quasinormal}. These modes are solutions to the perturbation equations that satisfy the boundary conditions of purely outgoing waves at infinity and purely ingoing waves at the event horizon \cite{kokkotas1999quasi,berti2009quasinormal}. In the context of nonlinear BH solutions in alternative theories of gravity, QNMs provide a powerful tool for probing the underlying structure of spacetime and the modifications introduced by these theories \cite{Stefanov:2007eq,Doneva:2010ke,Konoplya:2011qq,Konoplya:2013rxa,Kokkotas:2015uma,Blazquez-Salcedo:2019nwd,Daghigh:2020jyk,Okyay:2021nnh,Langlois:2021aji,Konoplya:2023ppx}. For instance, in theories like EEH or other nonlinear electrodynamics frameworks, the corrections to classical electrodynamics and GR can influence the QNM spectrum, leading to potentially observable deviations from the predictions of GR \cite{Breton:2021mju,Luo:2022gdz,Feng:2022otu}. Studying QNMs in such scenarios allows for testing the validity of alternative gravity theories and exploring the role of nonlinear effects, offering insights into high-energy physics and quantum gravitational phenomena. This makes QNMs a critical probe for connecting theoretical models with astrophysical observations, such as those from gravitational wave (GW) detectors \cite{Nakamura:2016gri,Aneesh:2018hlp,Yi:2024elj}.
This paper aims to explore the characteristics of a dyonic charged BH within the EEH theory, focusing on how this framework differs from the separate electric and magnetic charged BH in this theory. We also focus to analyze the impact of nonlinear electromagnetic corrections on key properties such as the oscillation, damping rate and grey-body factors, and discuss the potential observational consequences of these deviations.\\
The paper is structured as follows: Section \ref{sec2} provides a detailed explanation of the dyonic BH within the framework of EEH theory, elucidating its theoretical foundations and key characteristics. In Section \ref{sec3}, we analyze scalar perturbations and quasinormal modes. Section \ref{sec4} investigates the grey body factor. Finally, the conclusions are summarized in Section \ref{sec5}.
\section{Dyonic BH in EEH Theory} \label{sec2}
We consider the action of the Einstein general relativity theory minimally coupled to a NLED as \cite{Ruffini:2013hia},
\begin{equation}
    S = \frac{1}{4\pi G} \int_{M^{4}} d^{4} x \sqrt{-g} \left[ \frac{1}{4} R - \mathcal{L} (X,Y)\right],
\end{equation}
here, $R$ is the ricci scalar and $\mathcal{L} (X,Y)$ is the Lagrangian of the NLED
theory. This Lagrangian depend only on two independent relativistic invariants constructed from the Faraday tensor of the Maxwell field in four dimensions: the scalar $X$ and the pseudoscalar $Y$ defined as follows:
\begin{equation}
    X = \frac{1}{4} F_{\mu\nu} F^{\mu\nu} = \frac{1}{2} (B^{2}-E^{2}), \And
    Y = \frac{1}{4} F_{\mu\nu} {^*}F^{\mu\nu} = E. B.
\end{equation}
where E and B are the electric field and the magnetic field strength, respectively, while $F_{\mu\nu}$ is the Faraday electromagnetic tensor, and ${^*}F_{\mu\nu}$ is the dual ($*$ is Hodge dual operator) defined as,
\begin{equation*}
{^*}F_{\mu\nu} = \frac{1}{2} \sqrt{-g} \epsilon_{\mu\nu\sigma\rho} F^{\sigma\rho}, \hspace{0.3cm} \epsilon_{0123} =-1,
\end{equation*}
\begin{equation}
{^*}F^{\mu\nu} = \frac{1}{2} \frac{1}{\sqrt{-g}} \epsilon^{\mu\nu\sigma\rho} F_{\sigma\rho}, \hspace{0.3cm} \epsilon^{0123} =1,
\end{equation}
$\epsilon_{\mu\nu\sigma\rho}$ is completely antisymmetric and satisfies $\epsilon_{\mu\nu\sigma\rho}\epsilon^{\mu\nu\sigma\rho} = -4 !$.
For the case of NLED, and specifically for EEH theory, the detailed derivations can be found in \cite{Nomura:2020tpc,Magos:2023nnb}. We focus on the case of a dyonic charge, for which the electromagnetic gauge potential corresponding to electric and magnetic charges, $Q$ and $Q_{m}$ respectively, can be expressed as follows \cite{Hawking:1995ap},
\begin{equation}
    A_{\mu} = [A_{t}(r), 0, 0, -Q_{m}\cos{\theta}],
\end{equation}
while the nonzero components of the Faraday tensor $F_{\mu\nu}$, are $F_{01}$ = -$F_{10}$, for the electric field, and $F_{23}$ = -$F_{32}$, for the magnetic field. The t-component of the vector potential $A_{t}$ takes the form,
\begin{equation}
    A_{t}(r) = \frac{Q}{r} \left( 1- \frac{2\alpha}{225 \pi} E_{Q}^{2} - \frac{\alpha}{45\pi} B_{Q}^{2} \right),
\end{equation}
where $B_{Q}=Q_{m}/r^{2}E_{c}$ is the magnetic field strength, and the electric field $E_{Q}=Q/r^{2}E_{c}$, with $E_{c}=m^{2}c^{3}/eh$. $\alpha$ is the nonlinear electrodynamics correction parameter. The electric and magnetic field read,
\begin{equation}
    E(r) = \frac{Q}{r^{2}} -\frac{2\alpha}{45\pi}\frac{Q^{3}}{E_{c}^{2}r^{6}} - \frac{\alpha}{9\pi}\frac{B(r)^{2}}{E_{c}^{2}}\frac{Q}{r^{2}},
\end{equation}
with
\begin{equation}
    B(r) = \frac{Q_{m}}{r^{2}}.
\end{equation}
Therefore, the nonzero components of the Faraday tensor are given by,
\begin{equation}
    F_{01} = \frac{Q}{r^{2}} \left( 1- \frac{2\alpha}{45\pi} E_{Q}^{2} -\frac{\alpha}{9\pi} B_{Q}^{2} \right), \hspace{0.5cm} F_{23} = Q_{m} \sin{\theta}. 
\end{equation}
In this case, both $X$ and $Y$ invariants are non-vanishing and reads as,
\begin{equation}
    X = - \frac{Q}{2r^{4}} \left( 1- \frac{2\alpha}{45\pi} E_{Q}^{2} -\frac{\alpha}{9\pi} B_{Q}^{2} \right) + \frac{Q_{m}^{2}}{2r^{4}},
\end{equation}
\begin{equation}
    Y = - \frac{QQ_{m}}{r^{4}} \left( 1- \frac{2\alpha}{45\pi} E_{Q}^{2} -\frac{\alpha}{9\pi} B_{Q}^{2} \right).
\end{equation}
The $tt$-component of the Einstein field equations leads to,
\begin{equation}
    m'(r) = \frac{E_{c}^{2}r^{2}}{2} \left( E_{Q}^{2} + B_{Q}^{2} - \frac{\alpha}{45\pi}E_{Q}^{4} -\frac{\alpha}{45\pi}B_{Q}^{4} - \frac{\alpha}{9\pi} B_{Q}^{2} E_{Q}^{2} \right).
\end{equation}
With the framework established, we can consider a BH solution whose line element in spherical symmetric coordinates is given by,

\begin{equation}
    ds^2 = -f(r)dt^2 + \frac{1}{f(r)}dr^2 + r^2(d\theta^2 + \sin{\theta}^2d\phi^2),
\end{equation}
where the first order in $\alpha$, the metric function  of dyonic is defined as \cite{Magos:2023nnb},
\begin{equation}
f(r) = 1 - \frac{2M}{r} + \frac{\hat{Q}^2}{r^2} + \frac{\hat{Q}_m^2}{r^2},
\end{equation}
where, the screened electric and magnetic charges of the BH are given by,
\begin{equation}
    \hat{Q} = Q \left( 1 - \frac{\alpha E_Q^2}{225\pi} - \frac{\alpha B_Q^2}{90\pi }\right)^{\frac{1}{2}},
\end{equation}

\begin{equation}
    \hat{Q}_m = Q_m \left( 1 - \frac{\alpha B_Q^2}{225\pi} - \frac{\alpha E_Q^2}{90\pi} \right)^{\frac{1}{2}}.
\end{equation}
The dyonic BH solution in EEH theory reduces to the well-known Reissner-Nordström (RN) BH in the specific limits $\hat{Q}_{m}=0$ (no magnetic charge) and $\alpha=0$ (no nonlinear electrodynamics corrections). Additionally, the metric can be specialized to describe an electrically charged BH in EEH theory by setting $\hat{Q}_{m}=0$, or a magnetically charged BH in EEH theory by setting $\hat{Q}=0$.

\section{Scalar Perturbations and Quasinormal Modes } \label{sec3}

The basic equation for a neutral scalar field is given by the Klein-Gordon (KG) equation as follows,
\begin{equation}
    \frac{1}{\sqrt{-g}} \partial_{\nu} \left( g^{\mu\nu} \sqrt{-g} \partial_{\mu} \Phi_{s}\right) =0.
\end{equation}
Given the spherical symmetry of the geometry under investigation, we can employ the following spherical harmonics to separate the variables,
\begin{equation}
    \Phi (t, r, \theta, \phi) = \sum\limits_{l,m} Y_{l,m} (\theta, \phi) \frac{\psi_{l,m} (t,r)}{r},
\end{equation}
where, $Y_{l,m} (\theta, \phi)$ is the spherical harmonics. Here, $l$ and $m$ represent angular and azimuthal quantum numbers, respectively. For a given $l$ and $m$, we have simplified the notation by denoting $\psi_{l,m} (t,r)$ as $\psi$. Once separated, the radial part the KG equation reads as,
\begin{equation}
    \frac{d^2\psi_2}{dr^2_*} + (\omega^2 - V_s(r))\psi_s = 0,
\end{equation}
where $s=0$ corresponds to scalar field and $s=1$ to electromagnetic field. Here, $r_{*}$ is the tortoise coordinate associated with $dr_* = \frac{dr}{f(r)}$ and given as,
\begin{equation}
    dr_* = \frac{dr}{1 - \frac{2M}{r} + \frac{\hat{Q}^2}{r^2} + \frac{\hat{Q}^2_m}{r^2}}.
\end{equation}
Therefore, the scalar effective potential is given as follows,
\begin{equation}
    V_{scalar}(r_{*}) = f(r)\left( \frac{l(l+1)}{r^2} + \frac{1}{r}\frac{df(r)}{dr} \right).
\end{equation}
\begin{figure}[H]
    \centering
    \includegraphics[width=.6\linewidth,height=0.35\textheight]{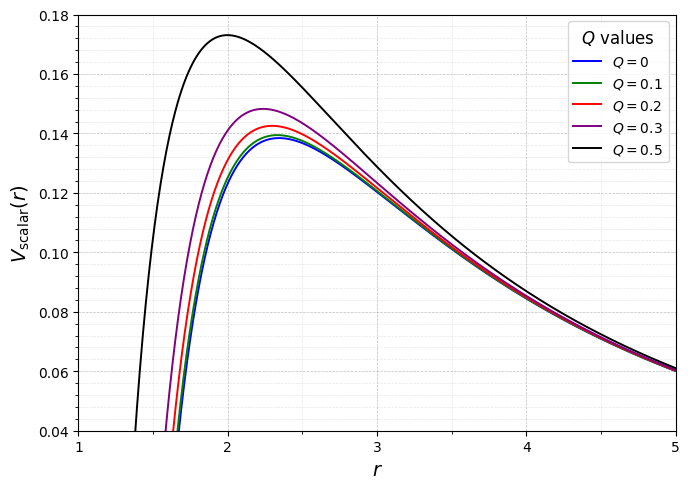}
    \caption{ Effective potential of dyonic BH in scalar field as a function of radial distance for different values of $Q$. Here we consider $l=1$ and $Q_m = 0.8$.}
    \label{fig:enter-label1}
\end{figure}

\begin{figure}[H]
    \centering
    \begin{minipage}{0.48\textwidth}
        \centering
        \includegraphics[width=\linewidth,height=0.35\textheight]{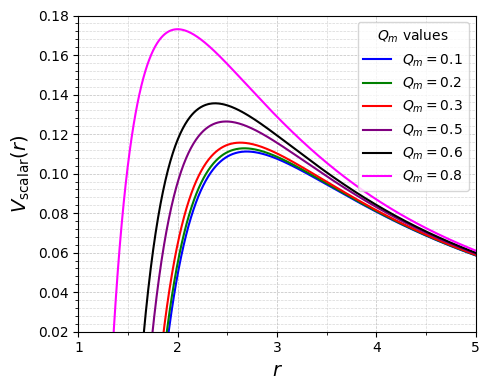}
        \caption{Effective potential of dyonic BH in the scalar field as a function of radial distance for different values of \( Q_{m} \). Here, we consider \( l=1 \) and \( Q = 0.5 \).}
        \label{fig:enter-label2}
    \end{minipage}
    \hfill
    \begin{minipage}{0.48\textwidth}
        \centering
        \includegraphics[width=\linewidth,height=0.35\textheight]{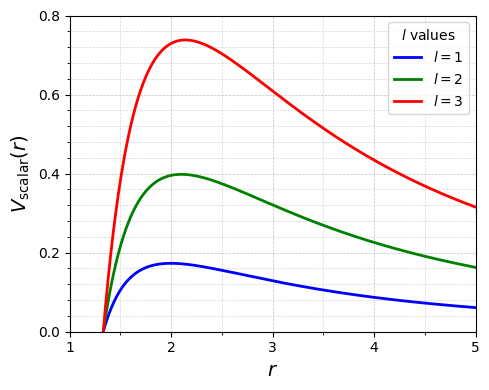}
        \caption{Effective potential of dyonic BH in the scalar field as a function of radial distance for different values of \( l \). Here, we consider \( Q=0.5 \) and \( Q_{m}=0.8 \).}
        \label{fig:enter-label3}
    \end{minipage}
\end{figure}
The graphical representations of the scalar potential for electrically charged and magnetically charged BHs in EEH theory are shown in Fig. \ref{fig:enter-label1} and Fig. \ref{fig:enter-label2}, respectively. The scalar potential is plotted for various values of the charge parameters $Q$ and $Q_m$, revealing that the potential increases as both charge parameters increase. In Fig. \ref{fig:enter-label3}, the scalar potential for a dyonic BH in EEH theory is displayed for different values of angular momentum. The scalar potential is observed to exhibit a minimum at the lowest value of $l$.\\
The method of using the Wentzel–Kramers–Brillouin (WKB) approximation to solve QNMs was first proposed by Schutz and Will \cite{schutz1985black}. This approach is particularly effective for calculating effective potentials that exhibit potential barriers and exhibit constant behavior near the boundaries. Later, Iyer and Will \cite{Iyer:1986np} extended this method to the third-order WKB approximation. Konoplya \cite{Konoplya:2003ii} further improved this technique, extending it to the sixth-order WKB approximation, and applied it to calculate the QNMs of D-dimensional Schwarzschild BHs. More recently, Matyjasek and Opala \cite{Matyjasek:2017psv} combined the Pade approximation with the WKB method, improving the accuracy to the 13th order. Using this framework, QNMs can be consistently described within the WKB approximation approach as given,
\begin{figure}[H]
    \centering
    \begin{minipage}{0.48\textwidth}
        \centering
        \includegraphics[width=\linewidth,height=0.35\textheight]{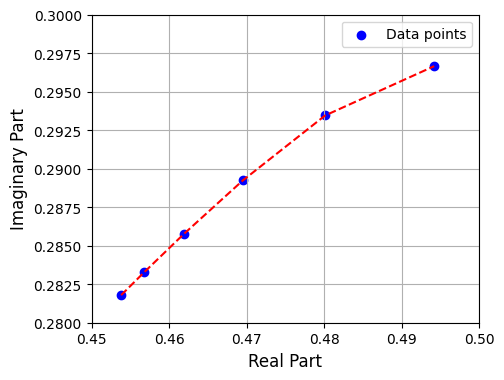}
        \caption{Variation of QNMs of dyonic BH with \( l=0 \) for scalar field.}
        \label{fig:enter-label5}
    \end{minipage}
    \hfill
    \begin{minipage}{0.48\textwidth}
        \centering
        \includegraphics[width=\linewidth,height=0.35\textheight]{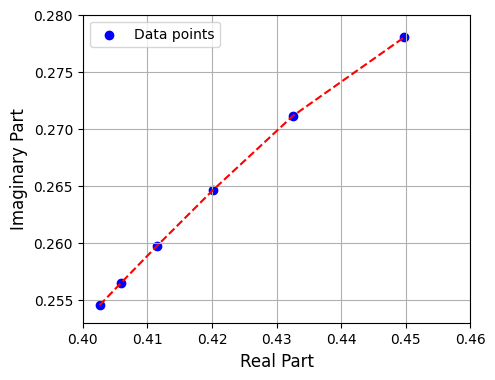}
        \caption{Variation of QNMs of dyonic BH with \( l=1 \) for electromagnetic field.}
        \label{fig:enter-label6}
    \end{minipage}
\end{figure}
\begin{equation} \label{Eq21}
\omega^2 = V_{max} - i \left( n + \frac{1}{2} + \sum\limits_{i=2}^{6} \bar{\Gamma_{k}} \right) \sqrt{-2V^{''}_{max}}.
\end{equation}
Here, $V_{max}$ is the maximum value of potential and $V_{max}^{''}$ is the second derivative of
the potential at the peak of the potential. The correction term $\bar{\Gamma_{k}}$ are crucial for enhancing the accuracy of the calculations and detailed expression can be found in here \cite{Konoplya:2003ii}. Here, $n$ represents the mode number. Numerous attempts have been made to derive specific expressions for these correction terms, often in conjunction with the Pade averaging method. References to these efforts can be found here \cite{Konoplya:2019hlu,Gogoi:2021cbp,Gogoi:2021dkr,konoplya2023quasinormal,Gogoi:2023fow,Gogoi:2023lvw,Gogoi:2024epx}. Since $n=0$ corresponds to the lowest frequency, which results in the longest time for the mode to decay, the stability of the BH is primarily determined by the fundamental frequency. Therefore, our primary focus will be on the fundamental frequency, corresponding to $n=0$. The WKB method is employed in this study as it offers a reliable semi-analytical approach for computing QNMs, particularly for higher angular momentum modes. Its balance between accuracy and computational efficiency makes it well-suited for analyzing BH perturbations. Moreover, this method has been extensively validated across various spacetimes, further supporting its applicability to our study.\\
In Fig. \ref{fig:enter-label5}, we present the real and imaginary parts of the QNM frequencies for scalar perturbations with $l=0$. Our analysis reveals that in the absence of angular momentum, the oscillations exhibit rapid damping. A similar trend is observed for electromagnetic perturbations, as shown in Fig. \ref{fig:enter-label6}. However, we find that the electromagnetic case follows a more linear behavior compared to the scalar case. This suggests that electromagnetic perturbations exhibit a more uniform dependence, indicating a weaker sensitivity to the BH’s electric and magnetic charge parameters.

\begin{figure}[H]
    \centering
    \includegraphics[width=.65\linewidth,height=0.65\textheight]{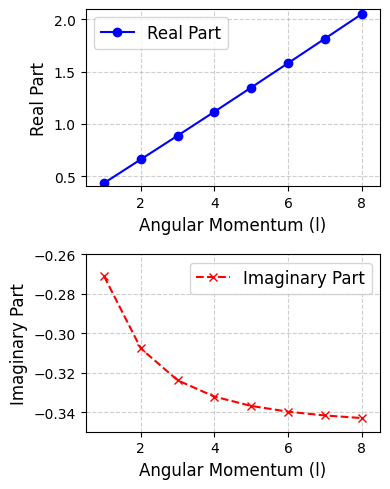}
    \caption{Variation of real and imaginary parts of QNMs of dyonic BH as a function of angular momentum with $Q_m=0.8$ and $Q=0.5$ for electromagnetic field.}
    \label{fig:enter-label7}
\end{figure}
\begin{figure}[H]
    \centering
    \begin{minipage}{0.48\textwidth}
        \centering
        \includegraphics[width=\linewidth,height=0.75\textheight]{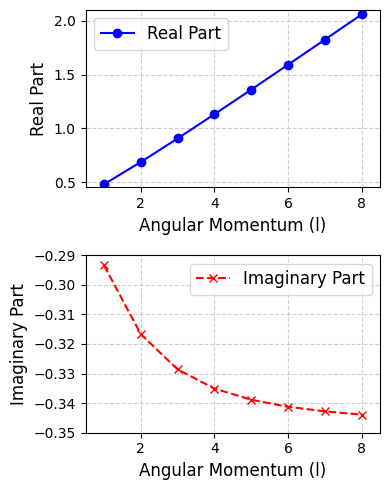}
        \caption{Variation of real and imaginary parts of QNMs of dyonic BH as a function of angular momentum with \( Q_m=0.8 \) and \( Q=0.5 \) for scalar field.}
        \label{fig:enter-label8}
    \end{minipage}
    \hfill
    \begin{minipage}{0.48\textwidth}
        \centering
        \includegraphics[width=\linewidth,height=0.75\textheight]{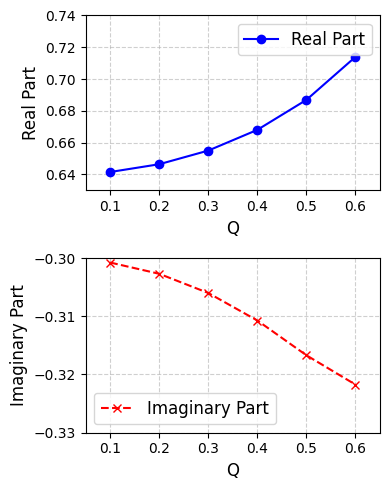}
        \caption{Variation of real and imaginary parts of QNMs of dyonic BH as a function of electric charge with \( Q_m=0.8 \) and \( l=2 \) for scalar field.}
        \label{fig:enter-label9}
    \end{minipage}
\end{figure}
\begin{figure}[H]
    \centering
    \begin{minipage}{0.48\textwidth}
        \centering
        \includegraphics[width=\linewidth,height=0.75\textheight]{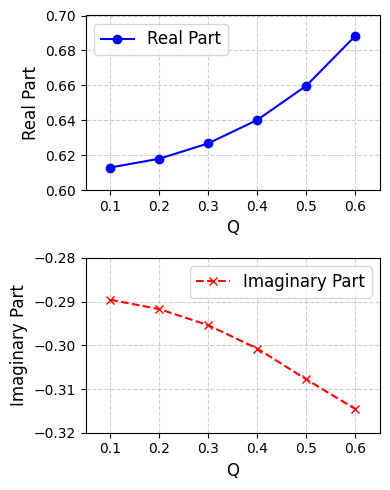}
        \caption{Variation of real and imaginary parts of QNMs of dyonic BH as a function of electric charge with \( Q_m=0.8 \) and \( l=2 \) for the electromagnetic field.}
        \label{fig:enter-label10}
    \end{minipage}
    \hfill
    \begin{minipage}{0.48\textwidth}
        \centering
        \includegraphics[width=\linewidth,height=0.75\textheight]{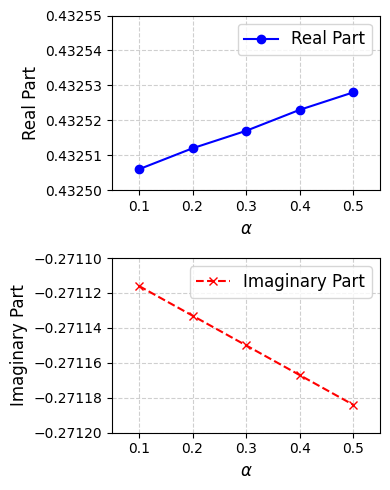}
        \caption{Variation of real and imaginary parts of QNMs of dyonic BH as a function of \( \alpha \) with \( Q_m=0.8 \) and \( Q=0.5 \) for the electromagnetic field.}
        \label{fig:enter-label12}
    \end{minipage}
\end{figure}

The graphical representation of QNMs of dyonic BH for electromagnetic and scalar potential considering $n=0$ are depicted in Fig. \ref{fig:enter-label7} and Fig. \ref{fig:enter-label8}, respectively. As anticipated, the real part of the QNM increases with the quantum number of angular momentum $l$, indicating higher oscillation frequencies for modes with larger values of $l$. The imaginary part, which governs the damping or decay rate, shows only a slight variation with $l$. Notably, scalar perturbations consistently exhibit slightly higher oscillation frequencies and more significant damping compared to electromagnetic fields. This is reflected in the larger imaginary components for scalar modes, in contrast to their electromagnetic counterparts, highlighting the distinct behaviors of these two types of perturbation in deformed AdS BH spacetimes. In Fig. \ref{fig:enter-label9} and Fig. \ref{fig:enter-label10}, we present the variation of QNMs with respect to the electric charge parameter for scalar and electromagnetic fields, respectively. In both cases, we observe that the oscillation frequency increases with the charge parameter, while the damping rate decreases as the charge parameter increases. In particular, scalar perturbations exhibit a slightly enhanced damping rate compared to electromagnetic fields, highlighting the distinct behavior between these two types of perturbations in the presence of an electric charge. Fig. \ref{fig:enter-label12} illustrates the impact of the nonlinear correction parameter ($\alpha$) on QNM frequencies. The real part increases linearly with $\alpha$, indicating a higher oscillation frequency, while the imaginary part decreases linearly, suggesting slower decay of perturbations. This trend implies that stronger nonlinear corrections enhance oscillations while reducing energy dissipation.

For numerical accuracy, we present the obtained results in tabular form. Table \ref{tab:1} displays the QNM frequencies for scalar and electromagnetic fields by varying the electric charge while keeping the magnetic charge fixed. Table \ref{tab:2} presents the QNMs for the cases where the magnetic charge is fixed and the electric charge is varied, and vice versa. Table \ref{tab:3} focuses on electromagnetic perturbations, where the electric charge is fixed, and the magnetic charge is varied. In Table \ref{tab:4}, we analyze the effect of the angular momentum parameter on QNMs while keeping both charges fixed. Finally, Table \ref{tab:5} illustrates the impact of different nonlinear parameters on QNMs.\\
\begin{table}[H]
\centering
\renewcommand{\arraystretch}{0.8} 
\setlength{\tabcolsep}{18pt} 
\caption{The QNMs of Scalar and Electromagnetic Field for a Dyonic BH (\( l=1 \), \( n=0 \), \( \alpha=0.1 \)), \( Q_m=0.8 \) and varying \( Q \)}
\label{tab:1}
\resizebox{\textwidth}{!}{ 
\begin{tabular}{cccc}
\toprule
\textbf{S.No.} & \textbf{\( Q \)} & \textbf{WKB (Scalar)} & \textbf{WKB (EM)} \\
\midrule
1  & 0.1  & \( 0.453822 - 0.281791i \) & \( 0.402610 - 0.254536i \) \\
2  & 0.2  & \( 0.456774 - 0.283267i \) & \( 0.405848 - 0.256449i \) \\
3  & 0.3  & \( 0.461895 - 0.285754i \) & \( 0.411524 - 0.259752i \) \\
4  & 0.4  & \( 0.469516 - 0.289236i \) & \( 0.420127 - 0.264609i \) \\
5  & 0.5  & \( 0.480150 - 0.293459i \) & \( 0.432506 - 0.271116i \) \\
6  & 0.6  & \( 0.494217 - 0.296660i \) & \( 0.449755 - 0.278016i \) \\
7  & 0.7  & \( 0.501942 - 0.272458i \) & \( 11.5121 - 6.50393i \) \\
8  & 0.8  & \( 1.354680 - 1.16046i \) & \( 12.0072 - 6.40464i \) \\
9  & 0.9  & \( 1.238210 - 1.01233i \) & \( 12.6246 - 6.34796i \) \\
10 & 1.0  & \( 1.146900 - 0.902665i \) & \( 13.3192 - 6.31495i \) \\
\bottomrule
\end{tabular}
}
\end{table}

\begin{table}[H]
\centering
\renewcommand{\arraystretch}{0.8} 
\setlength{\tabcolsep}{18pt} 
\caption{The QNMs of scalar field for a dyonic BH (\( l=0 \), \( n=0 \), \( \alpha=0.1 \)), $Q_{m}=0.8$ and varying $Q$ and $Q=0.5$ varying $Q_m$ }
\label{tab:2}
\resizebox{\textwidth}{!}{ 
\begin{tabular}{ccccc}
\toprule
\textbf{$S.No.$} & \textbf{$Q$} & \textbf{WKB} & \textbf{$Q_m$} & \textbf{WKB}  \ \\ 
\midrule
1.  & 0.1  & $0.31113 - 0.258638 i$ & 0.1  & $0.304329 - 0.254753i$ \\
2.  & 0.2  & $0.311335 - 0.258566i$ & 0.2  & $0.305002 - 0.255236i$ \\
3.  & 0.3  & $0.311489 - 0.258236i$ & 0.3  & $0.306094 - 0.256002i$ \\
4.  & 0.4  & $0.311221 - 0.25724i$ & 0.4  & $0.30754 - 0.256967i$ \\
5.  & 0.5  & $0.309814 - 0.254822i$ & 0.5  & $0.309199 - 0.257961i$ \\
6.  & 0.6  & $0.30621 - 0.250016i$ & 0.6  & $0.310755 - 0.258619i$ \\
7.  & 0.7  & $0.300953 - 0.244453i$ & 0.7  & $0.311494 - 0.25814i$ \\
8.  & 0.8  & $0.297428 - 0.244168i$ & 0.8  & $0.309814 - 0.254822i$ \\
9.  & 0.9  & $0.28835 - 0.240759i$ & 0.9  & $0.303648 - 0.247041i$ \\
10. & 1.0  & $0.270066 - 0.228348i$ & 1.0  & $0.298181 - 0.244033i$ \\
\bottomrule
\end{tabular}
}
\end{table}

\begin{table}[H]
\centering
\renewcommand{\arraystretch}{0.8} 
\setlength{\tabcolsep}{18pt} 
\caption{The QNMs of Scalar and electromagnetic field for a dyonic BH (\( l=1 \), \( n=0 \), \( \alpha=0.1 \)), $Q=0.5$ and varying $Q_m$}
\label{tab:3}
\resizebox{\textwidth}{!}{ 
\begin{tabular}{cccc}
\toprule
\textbf{$S.No.$} & \textbf{$Q_m$} & \textbf{WKB (Scalar)} & \textbf{WKB (EM)} \\
\midrule
1.  & 0.1  & $0.42185 - 0.264476 i$ & $0.368818 - 0.233824i$ \\
2.  & 0.2  & $0.423967 - 0.265677i$ & $0.370995 - 0.235187i$ \\
3.  & 0.3  & $0.423967 - 0.265677 i$ & $0.37475 - 0.237531i$ \\
4.  & 0.4  & $0.432934 - 0.270695i$ & $0.3803 - 0.240979i$ \\
5.  & 0.5  & $0.440253 - 0.274693 i$  & $0.388008 - 0.245725i$ \\
6.  & 0.6  & $0.450018 - 0.279852i$  & $0.398471 - 0.252069i$ \\
7.  & 0.7  & $0.462952 - 0.286256i$ & $0.412704 - 0.260432i$ \\
8.  & 0.8  & $0.48015 - 0.293459i$   & $0.432506 - 0.271116 i$ \\
9.  & 0.9  & $0.501474 - 0.294778i$   & $0.458833 - 0.277372i$ \\
10. &1.0  & $1.37674 - 1.1916i$   & $14.9229 - 7.34635i$ \\
\bottomrule
\end{tabular}
}
\end{table}

\begin{table}[H]
\centering
\renewcommand{\arraystretch}{0.8} 
\setlength{\tabcolsep}{18pt} 
\caption{The QNMs of Scalar and electromagnetic field for a dyonic BH (\( n=0 \), \( \alpha=0.1 \)), $Q=0.5$, $Q_m=0.8$ and varying $l$}
\label{tab:4}
\resizebox{\textwidth}{!}{ 
\begin{tabular}{cccc}
\toprule
\textbf{$S.No.$} & \textbf{$l$} & \textbf{WKB (Scalar)} & \textbf{WKB (EM)} \\
\midrule
1.  & 1  & $0.48015 - 0.293459 i$ & $0.432506 - 0.271116i$ \\
2.  & 2  & $0.68688 - 0.316704i$ & $0.659856 - 0.307793i$ \\
3.  & 3  & $0.905619 - 0.328607 i$ & $0.886719 - 0.32392i$ \\
4.  & 4  & $1.13078 - 0.335069i$ & $1.11621 - 0.3322 i$ \\
5.  & 5  & $1.35978 - 0.338852 i$  & $1.34792 - 0.336922i$ \\
\bottomrule
\end{tabular}
}
\end{table}

\begin{table}[H]
\centering
\renewcommand{\arraystretch}{0.8} 
\setlength{\tabcolsep}{18pt} 
\caption{The QNMs of Scalar and electromagnetic field for a dyonic BH (\(l=1\),\( n=0 \)), $Q=0.5$, $Q_m=0.8$ and varying $\alpha$}
\label{tab:5}
\resizebox{\textwidth}{!}{ 
\begin{tabular}{cccc}
\toprule
\textbf{$S.No.$} & \textbf{$\alpha$} & \textbf{WKB (Scalar)} & \textbf{WKB (EM)} \\
\midrule
1.  & 0.1  & $0.48015 - 0.293459 i$ & $0.432506 - 0.271116i$ \\
2.  & 0.2  & $0.48016 - 0.293479i$ & $0.432512 - 0.271133i$ \\
3.  & 0.3  & $0.48017 - 0.293499i$ & $0.432517 - 0.27115i$ \\
4.  & 0.4  & $0.480179 - 0.293519i$ & $0.432523 - 0.271167i$ \\
5.  & 0.5  & $0.480189 - 0.293539i$  & $0.432528 - 0.271184i$ \\
\bottomrule
\end{tabular}
}
\end{table} 
We compare the obtained results with the QNMs studied in EEH theory for purely electric and purely magnetic charges \cite{Breton:2021mju}, as well as with the classical RN BH results from Kokkotas et al., \cite{kokkotas1988black}. Our findings indicate that the presence of both electric and magnetic charges (dyonic case) introduces modifications to the QNMs spectrum that are not simply additive from the individual electric or magnetic cases. In comparison to the EEH BH with only electric or magnetic charge, we observe that dyonic BH exhibit a shift in both the real and imaginary parts of the QNMs. The real frequency tends to increase, indicating a stronger oscillatory nature, while the imaginary part shows deviations that influence the damping rate. This suggests that the dyonic configuration affects the stability and response of perturbations in contrast to purely charged EEH BHs. Compared with Kokkotas's RN results, the inclusion of nonlinear electrodynamics in EEH modifies QNMs by introducing additional dependencies on the nonlinear parameter and dyonic charge structure. The deviations highlight the role of nonlinear corrections in the electromagnetic field along with dyonic charge, which alters the ringdown properties.
\section{Greybody factor} \label{sec4}
\begin{figure}[H]
    \centering
    \begin{minipage}{0.48\textwidth}
        \centering
        \includegraphics[width=\linewidth,height=0.35\textheight]{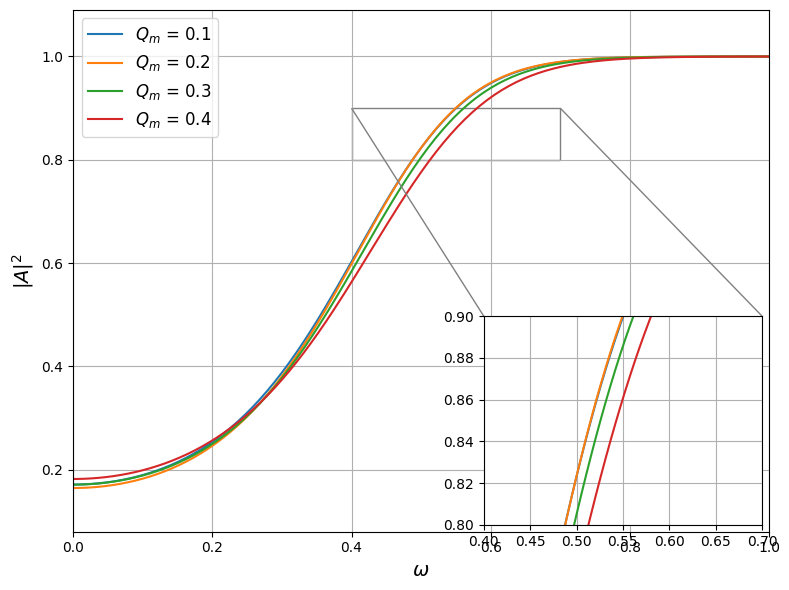}
        \caption{Greybody factor for dyonic BH in the scalar field, with \( Q=0.8 \), \( \alpha=0.1 \), and \( l=1 \).}
        \label{fig:enter-label14}
    \end{minipage}
    \hfill
    \begin{minipage}{0.48\textwidth}
        \centering
        \includegraphics[width=\linewidth,height=0.35\textheight]{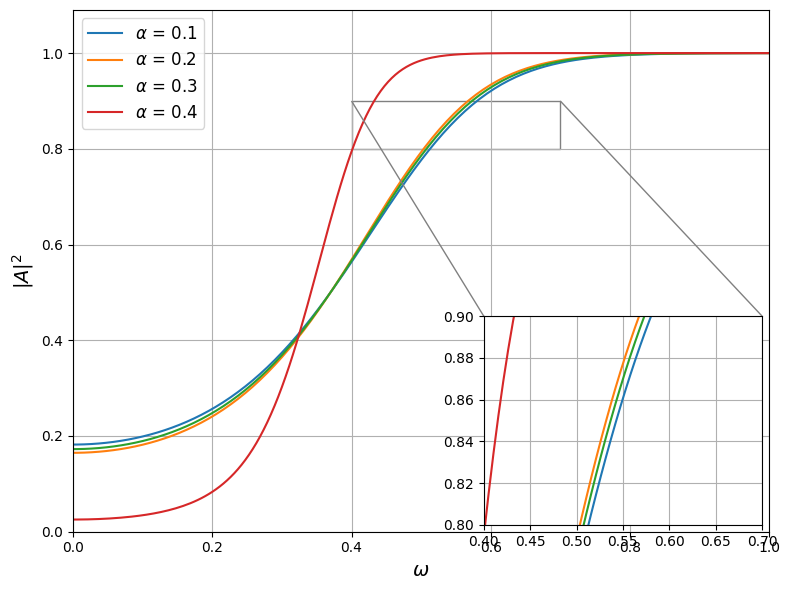}
        \caption{Greybody factor for dyonic BH in the scalar field, with \( Q=0.8 \), \( l=1 \), and \( Q_m=0.4 \).}
        \label{fig:enter-label15}
    \end{minipage}
\end{figure}

In this section, we introduce the scheme of using the WKB method to analyze the greybody factor, which can be used to further describe the intrinsic characteristics of the effective potential of background spacetime. We begin by analyzing the wave equation under boundary conditions that permit incoming waves from infinity. Due to the symmetry of the scattering process, this setup is equivalent to studying the scattering of a wave originating from the BH horizon. This approach is particularly useful for determining the fraction of waves that are reflected back towards the horizon by the potential barrier. The boundary conditions for this scenario are given by,
\begin{equation}
    \psi_{s} = Te^{-i\omega r_{*}}, \hspace{0.5cm} r_{*} \longrightarrow -\infty
\end{equation}
\begin{equation}
    \psi_{s} = e^{-i\omega r_{*}} + R e^{i\omega r_{*}}, \hspace{0.5cm} r_{*} \longrightarrow \infty
\end{equation}
where, $T$ is the transmission coefficient and $R$ is the reflection  coefficient and satisfy the conservation probability as,
  \begin{equation}
      |T|^2 + |R|^2 = 1.
  \end{equation}
The transmission coefficient $|T|$ which is equivalent to the greybody factor $|A|$ of the BH, is calculated using the WKB approach:
    \begin{equation}
      |A_l|^2 = 1 - |R_l|^2 = |T_l|^2,
  \end{equation}
  where, the reflection coefficient $R$ is given by,
   \begin{equation}
     R = \sqrt{(1 + e^{-2 i \pi K})}.
 \end{equation}
 The phase vector $K$ is determined by the equation,
 \begin{equation}
     K - i (\frac{(\omega^2 - V_{max})}{\sqrt{-2V^{''}_{max}}}) - \sum\limits_{i=2}^{6} \Gamma_{i} (K) = 0.
 \end{equation}
 Here $V_{max}$ is maximum of effective potential, $V^{''}_{max}$ is second derivative of the effective potential. Also, $V_{max}$ and $V^{''}_{max}$ are already defined below \ref{Eq21}.
  \begin{figure}[H]
    \centering
    \includegraphics[width=.6\linewidth,height=0.35\textheight]{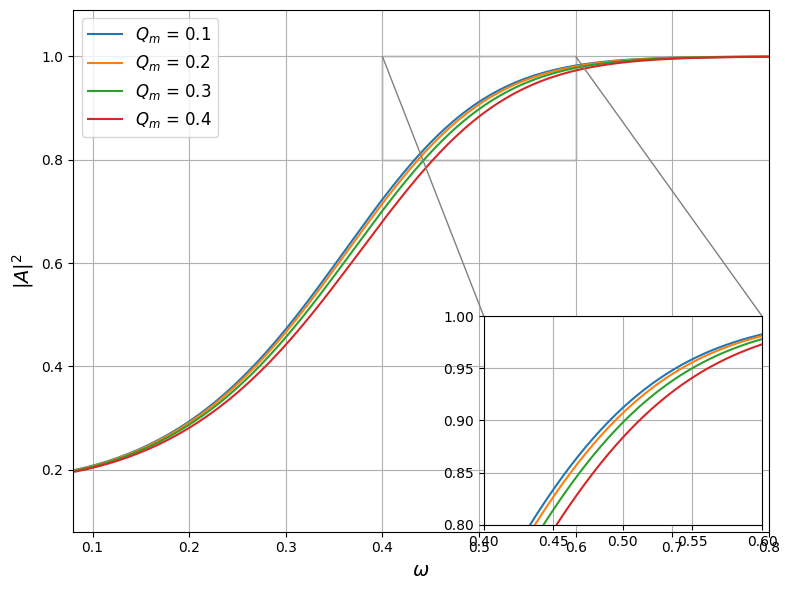}
    \caption{Greybody factor for dyonic BH in electromagnetic field, with $Q=0.8$, $\alpha=0.1$ and $l=1$.  }
    \label{fig:enter-label13}
\end{figure}
The graphical representation of the transmission probability as a function of specific frequencies are depicted to examine the influence of BH parameters. Figures \ref{fig:enter-label13} and \ref{fig:enter-label14} illustrate this behavior for the EM field and scalar field, respectively. In both cases, it is observed that the transmission probability decreases with an increase in charge parameter. Furthermore, beyond a specific frequency value, the transmission probability saturates, which is a typical phenomenon. The effect of the nonlinear electrodynamics correction on the greybody factor is depicted in Fig. \ref{fig:enter-label15}. The behavior of the transmission probability no longer remains constant for higher values of $\alpha$, exhibiting distinct variations. This indicates that the parameter $\alpha$ significantly influences the scattering and absorption processes, altering the effective potential experienced by the waves. Physically, this suggests that higher values of $\alpha$ modify the spacetime geometry around the BH, leading to changes in how waves propagate and interact with the BH's gravitational field. 

\section{Conclusions} \label{sec5}
\noindent In this study, we investigated the QNMs of a dyonic BH in EEH theory for both scalar and electromagnetic fields. The study explored the influence of key parameters, such as electric charge, magnetic charge, angular momentum, and the nonlinear dynamic correction parameter, on the QNMs frequencies.  Additionally, we compared the QNMs results obtained for the dyonic BH in EEH theory with those of the separate electric and magnetic charge BHs solution in EEH theory, as well as with the RN BH in GR. The main findings of this study are as follows:
\begin{itemize}
    \item Our findings reveal that the inclusion of both electric and magnetic charges in the EEH BH results in significant modifications to the QNMs compared to RN BH solution in GR. For scalar perturbations, the real part of the frequencies was observed to increase with the charges, while the imaginary part demonstrated a characteristic damping behavior influenced by the charge parameters.
    \item The presence of angular momentum introduces a shift in the QNMs, with higher angular momentum leading to an increase in the oscillation frequency and a decrease in the damping rate for both scalar and electromagnetic fields. Notably, scalar perturbations consistently exhibit slightly higher oscillation frequencies and more significant damping compared to electromagnetic fields.
    \item Further, we study the greybody factor and observed that the nonlinear correction parameter, introduced by the EEH theory, has a distinct impact on the QNM behavior, influencing both the oscillation frequencies and the damping rates, particularly for scalar perturbations. In comparison to electromagnetic fields, scalar modes exhibited more significant shifts in the QNMs spectra, with higher damping rates and lower oscillation frequencies in some cases. This suggests that scalar perturbations are more sensitive to the modifications introduced by the nonlinear electromagnetic interactions in the EEH framework.
    \item The results were further compared with those for the electric and magnetic charge BHs in EEH theory. Notably, the dyonic-charged EEH BH presented more pronounced modifications in the QNMs spectra, indicating the critical role of the nonlinear dynamics and the presence of both electric and magnetic charges in altering the BHs perturbative behavior.
\end{itemize}
The alterations in QNMs and damping rates, induced by the presence of dyonic and nonlinear electrodynamic corrections, imprint distinct signatures in the GWs spectrum of perturbed BHs. These deviations offer a potential observational avenue for testing the physics beyond GR, including the effects of nonlinear electrodynamics and alternative gravity theories, through future GWs detections.
\section*{Acknowledgments}
The authors acknowledge the use of facilities at ICARD, Gurukula Kangri (Deemed to be University), Haridwar, India. The author, SK, sincerely acknowledges IMSc for providing exceptional research facilities and a conducive environment that facilitated his work as an Institute Postdoctoral Fellow. One of the authors, H.N., would also like to thank IUCAA, Pune, for the support under its associateship program where a part of this work was done. The author HN also acknowledges the financial support provided by the Science and Engineering Research Board (SERB), New Delhi, through grant number CRG/2023/008980.

\bibliography{EEH}

\end{document}